\documentclass{INTERSPEECH2023}

\usepackage{algorithmic}
\usepackage{algorithm}
\usepackage{pdfpages}
\usepackage{bm}
\usepackage{amsmath, amssymb}
\usepackage{type1cm}
\usepackage{amssymb}
\usepackage{comment}
\usepackage{multirow}
\usepackage{cite}

\interspeechcameraready

\title{TENSOR DECOMPOSITION FOR MINIMIZATION OF E2E SLU MODEL \\TOWARD ON-DEVICE PROCESSING}
\name{
\begin{tabular}{c}
Yosuke Kashiwagi$^1$, Siddhant Arora$^2$, Hayato Futami$^1$, Jessica Huynh$^2$, Shih-Lun Wu$^2$, \\ Yifan Peng$^2$, Brian Yan$^2$, Emiru Tsunoo$^1$, Shinji Watanabe$^2$
\end{tabular}
}
\address{
  $^1$ Sony Group Corporation, Japan,\\
  $^2$ Carnegie Mellon University, USA}
\email{yosuke.kashiwagi@sony.com}

\begin{document}

\maketitle
 
\begin{abstract}
Spoken Language Understanding (SLU) is a critical speech recognition application and is often deployed on edge devices. 
Consequently, on-device processing plays a significant role in the practical implementation of SLU.
This paper focuses on the end-to-end (E2E) SLU model due to its small latency property, unlike a cascade system, and aims to minimize the computational cost.
We reduce the model size by applying tensor decomposition to the Conformer and E-Branchformer architectures used in our E2E SLU models.
We propose to apply singular value decomposition to linear layers and the Tucker decomposition to convolution layers, respectively.
We also compare COMP/PARFAC decomposition and Tensor-Train decomposition to the Tucker decomposition.
Since the E2E model is represented by a single neural network, our tensor decomposition can flexibly control the number of parameters without changing feature dimensions.
On the STOP dataset, we achieved 70.9\% exact match accuracy under the tight constraint of only 15 million parameters.
%Even though our model has a simple E2E structure, we obtain better results than previous studies under the same constraint.
\begin{comment}
On-device processing occupies a significant position in the practical application of neural nets.
In this paper, we describe our team's study on track 2 of the Spoken Language Understanding Grand Challenge, which is a part of ICASSP Signal Processing Grand Challenge 2023.
The task is designed for on-device processing and estimates semantic parse labels from speech with 15 million parameters.
We use an end-to-end E-Branchformer based spoken language understanding model, which is easier to control the number of parameters than cascade models, and reduced the parameter size by sequential distillation and tensor decomposition techniques.
On the STOP dataset, we achieved 70.9\% exact match under the tight constraint of 15 million parameters.
Even though our model has a simple E2E structure, we obtain better results than previous studies under the same constraint.
\end{comment}
\end{abstract}
%
%\begin{keywords}
%STOP challenge, spoken language understanding, E2E, on-device, sequential distillation, tensor decomposition, E-Branchformer
%\end{keywords}


\section{Introduction}
Spoken language understanding (SLU) is one of the essential applications of speech recognition.
SLU is used in voice interfaces such as smart speakers, and improving its accuracy is required for usability \cite{serdyuk2018towards,radfar2020end,dong2022improving,wang2022adversarial,arora2022two}.
% \cite{serdyuk2018towards,radfar2020end,qin2021survey,shon2022slue,cattan2022benchmarking,deng2022blockwise,dong2022improving,wang2022adversarial,arora2022two}
On the other hand, these voice interfaces often work on edge devices due to latency and privacy issues.
Such on-device processing requires smaller models to preserve power consumption.
Therefore, miniaturization of the SLU model for on-device processing is a critical issue for voice interfaces \cite{radfar2021fans,le2022deliberation,saade2019spoken,tyagi2020fast,desot2022end,gupta-2022-building,wang2022bottleneck,avila2022low,gao2022meta,dinarelli2022toward}.
% ,meeus2022multitask
% 先行研究もうちょっと増やす

For example, Radfar et al. reduce parameters by sharing the audio encoder and estimate slot tags, values, and intents \cite{radfar2020end,radfar2021fans}.
Also, Le et al. model NLU efficiently based on RNN-T using embedded features of both the encoder and the predictor of RNN-T \cite{le2022deliberation}.
In addition, Tyagi et al. propose early decision-making using the BranchyNet scheme to address the latency and computational complexity issues \cite{tyagi2020fast}. 
Although these are practical approaches, many are based on the traditional, relatively simple E2E model structure.
On the other hand, recent E2E models have been used with more complex structures by combining convolution and self-attention operations such as Conformer \cite{gulati2020conformer} and E-Branchformer \cite{peng2022branchformer,kim2023branchformer}.
%And E-Branchformer, a high-performance model with better parameter efficiency than the Conformer, has also been proposed \cite{kim2023branchformer}
Therefore, in addition to using singular value decomposition (SVD) \cite{michael2021comparison,zhang2021tiny,denton2014exploiting} for two-dimensional matrices in the self-attention network, 
%this paper proposes to combine tensor decomposition techniques for higher-order tensors used in the convolution network.
convolution network having high-order tensors requires tensor decomposition techniques.
We specifically target the Conformer and E-Branchformer-based E2E SLU model, which has demonstrated high accuracy in the SLU task\cite{arora2022espnet,peng2022branchformer}.
%Therefore, it is necessary to investigate decomposition methods not only for simple SVD targeting 2-D matrices but also for higher-order tensors.

%In this paper, we apply tensor decomposition techniques to the E2E SLU model.
%We propose using Conformer and E-Branchformer-based E2E SLU model.
%E-Branchformer is an enhanced version of Branchformer that achieves high performance in SLU tasks \cite{peng2022branchformer,kim2023branchformer}.
Tensor decomposition techniques are widely used for model miniaturization \cite{tjandra2017compressing}.
Although various tensor decomposition techniques have been proposed, we mainly explore SVD, and Tucker decomposition \cite{tucker1966some} to target on-device fast processing.
They enable inference without reconstructing the original parameter tensors from the factored tensors.
Decomposition with smaller ranks can then be used to reduce computational complexity, enabling a reduction in the number of parameters and faster computation \cite{kim2015compression}.
The paper demonstrates effective combinations of model compression techniques for Conformer and E-Branchformer, specifically through the use of tensor decompositions such as Tucker decomposition. 
This is because these models have convolution layers with higher-order tensors as parameters, making them more efficiently compressed using these techniques.
By defining the model compression ratio, we show that the model size can be flexibly changed by determining the rank of the decomposition from this ratio.
Furthermore, we evaluate CANDECOMP/PARAFAC (CP) decomposition \cite{harshman1970foundations} and Tensor-Train decomposition \cite{oseledets2011tensor} when used instead of the Tucker decomposition.

% It has been reported that Branchformer does not provide significant performance gains over conformer when the number of parameters is small \cite{peng2022branchformer}.
% We expect that model compression using tensor decomposition from a large E-Branchformer, rather than training from scratch, will mitigate the drawback.

\begin{comment}
This paper is based on our system submitted to the ICASSP Grand Challenge \footnote{https://facebookresearch.github.io/spoken\_task\_oriented\_parsing/}.
Spoken Language Understanding Grand Challenge, which is a part of ICASSP Signal Processing Grand Challenge 2023, is a competition to do semantic parsing from speech.
The challenge adopts STOP dataset \cite{paden2023stop}, which contains speech and semantic parse label pair data.
The challenge has three tracks.
Track 1 is an unrestricted track without closed data and models.
Track 2 is intended to operate on-device and is limited to a total parameter size within 15 million.
Track 3 is a low-resource track, and model adaptation is assumed.
This paper describes in detail our team's contribution to Track 2.
As an additional experiment, we investigate the performances with 30M parameter limitation as in the previous study \cite{le2022deliberation}, as well as with varying model compression ratios.
Furthermore, we also evaluate CANDECOMP/PARAFAC (CP) decomposition \cite{harshman1970foundations} and Tensor-Train decomposition \cite{oseledets2011tensor} when used instead of the Tucker decomposition.
\end{comment}
% We use sequential distillation and tensor decomposition techniques to Branchformer.

\section{Tensor decomposition}

\subsection{Singular value decomposition}
\label{sec:svd}
The SVD-based parameter reduction technique is stable and widely used \cite{michael2021comparison,zhang2021tiny,denton2014exploiting}.
The SVD is applied to the weight matrix $\bm{W} \in \mathbb{R}^{I\times J}$ with low-rank bases as:
%The weight matrix of the linear transformation $\bm{W} \in \mathbb{R}^{I\times J}$ is applied to the SVD with low-rank bases as:
\begin{align}
\bm{W} = \bm{U}\bm{S}\bm{V},
\end{align}
where $\bm{S}$ is a diagonal matrix.
The matrix size of each $\bm{U}$, $\bm{S}$ and $\bm{V}$ is $I\times R$, $R\times R$ and $R\times J$, respectively.
The parameter size can be controlled by changing $R$.
Since $\bm{S}$ is a square matrix, $\bm{U}\bm{S}$ or $\bm{S}\bm{V}$ can be pre-composed to reduce the parameters.
In our model, we composed $\bm{S}\bm{V}$.
Thus, the final number of decomposed parameters can be reduced into $IR + RJ$.
Therefore, the parameter compression ratio $\gamma_{\mbox{svd}}$ can be described as:
\begin{align}
\label{eq:svd_comp}
\gamma_{\mbox{svd}} = \frac{IR + RJ}{IJ}.
\end{align}

% \subsection{CP decomposition}

%\begin{comment}
%\begin{figure}[!t]
\begin{algorithm}[t]
    \caption{Size halving for Tucker decomposition}
    \label{alg1}
    \begin{algorithmic}[1]
    \REQUIRE $\hat{\gamma}_{\mbox{tucker}} > 0.0 $
    \STATE $R \leftarrow I, S \leftarrow J, T \leftarrow K$
    \STATE $\gamma_{\mbox{tucker}} \leftarrow \infty$
    \WHILE{$\gamma_{\mbox{tucker}} > \hat{\gamma}_{\mbox{tucker}} $}
    \STATE $R \leftarrow R / 2$, $S \leftarrow S / 2$, $T \leftarrow T / 2$
    \IF{ $R < 1$}
    \STATE $R \leftarrow 1$
    \ENDIF
    \IF{ $S < 1$}
    \STATE $S \leftarrow 1$
    \ENDIF
    \IF{ $T < 1$}
    \STATE $T \leftarrow 1$
    \ENDIF
    \STATE $\gamma_{\mbox{tucker}} \leftarrow \frac{ RST + IR + JS + KT }{ IJK }$
    \ENDWHILE
    \end{algorithmic}
\end{algorithm}
%\end{figure}
%\end{comment}

\subsection{Tucker decomposition}
\label{sec:tucker}
Tucker decomposition \cite{tucker1966some} is known to be effective for applying high-order tensors.
Here we discuss the case of 1d convolution, which has a 3-dimensional parameter tensor.
The parameter tensor of the convolution $\bm{W} \in \mathbb{R}^{I\times J\times K}$ can be decomposed into a core tensor $\bm{C} \in \mathbb{R}^{R\times S\times T}$ and factor matrices $\bm{U}^1 \in \mathbb{R}^{I\times R}$, $\bm{U}^2 \in \mathbb{R}^{J\times S}$ and $\bm{U}^3 \in \mathbb{R}^{K\times T}$ using Tucker decomposition as:
\begin{align}
\label{eq:tucker}
\bm{W} = \sum_{r,s,t} \bm{C}_{r,s,t} \times \bm{U}^1_{r} \otimes \bm{U}^2_{s} \otimes \bm{U}^3_{t},
\end{align}
where $\otimes$ describes a Kronecker product.
The parameter compression ratio $\gamma_{\mbox{tucker}}$ 
%of the Tucker decomposition
can be described as:
\begin{align}
\label{eq:tucker_comp}
\gamma_{\mbox{tucker}} = \frac{RST + IR + JS + KT}{IJK}.
\end{align}

%In our model, the Tucker decomposition is mainly used to reduce the parameters of the convolution layer.
%The reasons for adopting the Tucker decomposition are that it is expected to be able to represent the convolution weight tensor with fewer parameters than CP decomposition and that it contributes not only to parameter reduction but also to speeding up \cite{kim2015compression}.
%The tucker decomposition can be inferred without reconstructing the original tensor by swapping the order of computation as in SVD.

%For comparison, we use CANDECOMP/PARAFAC (CP) decomposition technique.

\subsection{CP decomposition}
\label{sec:cp}
CP decomposition \cite{harshman1970foundations} decomposes tensor $\bm{W} \in \mathbb{R}^{I\times J\times K}$ into a weight vector $\lambda \in \mathbb{R}^{R}$ and factor matrices $\bm{U}^1 \in \mathbb{R}^{I\times R}$, $\bm{U}^2 \in \mathbb{R}^{J\times R}$ and $\bm{U}^3 \in \mathbb{R}^{K\times R}$ as:
\begin{align}
\bm{W} = \sum_{r} \lambda_r \times \bm{U}^1_{r} \otimes \bm{U}^2_{r} \otimes \bm{U}^3_{r}.
\end{align}
The parameter compression ratio $\gamma_{\mbox{cp}}$ 
%of the Tucker decomposition
can be described as:
\begin{align}
\label{eq:cp_comp}
\gamma_{\mbox{cp}} = \frac{R(1 + I + J + K) }{IJK}.
\end{align}
The CP decomposition is the special case where the core tensor of the Tucker decomposition has only diagonal component values and the same rank for each dimension in eq. (\ref{eq:tucker}).
Therefore, it allows for more parameter reduction but is less expressive than the Tucker decomposition.
Furthermore, the rank of the CP decomposition is restricted to the smallest dimension.
Since the kernel size of the convolution layer is often small compared to the number of channels, the rank is often limited to the kernel size.
Therefore, CP decomposition is difficult to adjust with flexible parameter sizes.

\begin{comment}
\begin{table*}[t]
 \caption{Experimental results on STOP dataset within 15M parameters using Tucker decomposition.}
 \label{table:results_15M}
 \centering
  \begin{tabular}{lllccrc}
   \hline
    & & & Compression ratio & \multirow{2}{*}{Distillation} & \multirow{2}{*}{\# Parameters} & \multirow{2}{*}{Valid/Test EM} \\
    & & & encoder / decoder &  &  &  \\
   \hline \hline
   & Deliberation SLU \cite{le2022deliberation} & & &  & $< 15\mbox{M}$ & ----- / 67.9\\
   \hline
   \multirow{7}{*}{E2E} & \multirow{4}{*}{Conformer} & Large w/ Hubert & & & 430,635,985 & 68.8 / 69.4 \\
   & & Scratch & & \checkmark & 14,457,448 & 34.7 / 34.8 \\
   & & Middle & & \checkmark &  47,443,048 & 62.6 / 62.5 \\
   & & Small & 0.300 / 0.295 & \checkmark &  14,996,990 & 65.0 / 65.4 \\
   \cline{2-7}
   & \multirow{3}{*}{E-Branchformer} & Scratch & & \checkmark &  14,995,178 & 60.9 / 61.4 \\
   & & Middle & & \checkmark & 53,048,376 & 68.6 / 69.0\\
   & & Small  & 0.250 / 0.300 & \checkmark & 14,995,178 & {\bf 69.6 / 70.1 }\\
   \hline
  \end{tabular}
\end{table*}
\end{comment}

\begin{table*}[t]
 \caption{Experimental results on STOP dataset within 15M parameters using Tucker decomposition.}
 \label{table:results_15M}
 \centering
  \begin{tabular}{lllccrrc}
   \hline
    & & & Compression ratio & \multirow{2}{*}{Distillation} & \multirow{2}{*}{\# Epochs} & \multirow{2}{*}{\# Parameters} & \multirow{2}{*}{Valid/Test EM} \\
    & & & encoder / decoder &  & &  &  \\
   \hline \hline
   & Deliberation SLU \cite{le2022deliberation} & & & & & $< 15\mbox{M}$ & ----- / 67.9\\
   \hline
   \multirow{8}{*}{E2E} & \multirow{4}{*}{Conformer} & Teacher & & & & 431M & 68.8 / 69.4 \\
   & & Scratch & & \checkmark & 50 & 14M & 34.7 / 34.8 \\
   & & Middle & & \checkmark & 50 & 47M & 62.6 / 62.5 \\
   & & Small & 0.300 / 0.295 & \checkmark & 50 & 15M & 65.0 / 65.4 \\
   \cline{2-8}
   & \multirow{4}{*}{E-Branchformer} & Scratch & & \checkmark & 50 &  15M & 60.9 / 61.4 \\
   & & Middle & & \checkmark & 50 & 53M & 68.6 / 69.0\\
   & & Small  & 0.250 / 0.300 & \checkmark & 50 & 15M & 69.6 / 70.1\\
   & & Small  & 0.250 / 0.300 & \checkmark & 200 & 15M & {\bf 70.4 / 70.9 }\\
   \hline
  \end{tabular}
\end{table*}

\begin{comment}
\begin{table*}[t]
 \caption{Experimental results on STOP dataset within 30M parameters using Tucker decomposition.}
 \label{table:results_30M}
 \centering
  \begin{tabular}{lllccrc}
   \hline
    & & & Compression ratio & \multirow{2}{*}{Distillation} & \multirow{2}{*}{\# Parameters} & \multirow{2}{*}{Valid/Test EM} \\
    & & & encoder / decoder &  &  &  \\
   \hline \hline
   & Deliberation SLU \cite{le2022deliberation} & & &  & $< 30\mbox{M}$ & ----- / {\bf 73.87 }\\
   \hline
   \multirow{7}{*}{E2E} & \multirow{4}{*}{Conformer} & Large w/ Hubert & & & 430,635,985 & 68.8 / 69.4 \\
   & & Scratch & & \checkmark & 28,858,992 & 45.3 / 45.7 \\
   & & Middle & & \checkmark &  47,443,048 & 62.6 / 62.5 \\
   & & Small & 0.650 / 0.650 & \checkmark & 29,721,146  & 65.4 / 65.7 \\
   \cline{2-7}
   & \multirow{3}{*}{E-Branchformer} & Scratch & & \checkmark & 29,832,518 & 66.4 / 66.8\\
    & & Middle & & \checkmark & 53,048,376 & 68.6 / 69.0\\
   & & Small  & 0.550 / 0.600 & \checkmark & 29,832,518 & {\bf 71.0 } / 71.4\\
   \hline
  \end{tabular}
\end{table*}
\end{comment}

\begin{table*}[t]
 \caption{Experimental results on STOP dataset within 30M parameters using Tucker decomposition.}
 \label{table:results_30M}
 \centering
  \begin{tabular}{lllccrrc}
   \hline
    & & & Compression ratio & \multirow{2}{*}{Distillation} & \multirow{2}{*}{\# Epochs} & \multirow{2}{*}{\# Parameters} & \multirow{2}{*}{Valid/Test EM} \\
    & & & encoder / decoder &  &  &  \\
   \hline \hline
   & Deliberation SLU \cite{le2022deliberation} & & & & & $< 30\mbox{M}$ & ----- / {\bf 73.87 }\\
   \hline
   \multirow{7}{*}{E2E} & \multirow{4}{*}{Conformer} & Teacher & & & & 431M & 68.8 / 69.4 \\
   & & Scratch & & \checkmark & 50 & 29M & 45.3 / 45.7 \\
   & & Middle & & \checkmark & 50 & 47M & 62.6 / 62.5 \\
   & & Small & 0.650 / 0.650 & \checkmark & 50 & 30M  & 65.4 / 65.7 \\
   \cline{2-8}
   & \multirow{3}{*}{E-Branchformer} & Scratch & & \checkmark & 50 & 30M & 66.4 / 66.8\\
    & & Middle & & \checkmark & 50 & 53M & 68.6 / 69.0\\
   & & Small  & 0.550 / 0.600 & \checkmark & 50 & 30M & {\bf 71.0 } / 71.4\\
   \hline
  \end{tabular}
  \vspace{-8pt}
\end{table*}

\subsection{Tensor-Train decomposition}
\label{sec:tt}
Tensor-Train \cite{oseledets2011tensor} is another decomposition method for high-order tensor.
%We also use Tensor-Train decomposition for comparison.
The original tensor $\bm{W} \in \mathbb{R}^{I\times J\times K}$ is decomposed into a product of 3-order tensors $\bm{G}^1_{1,i,r} \in \mathbb{R}^{1\times I\times R}$, $\bm{G}^2_{r,j,s} \in \mathbb{R}^{R\times J\times S}$ and $\bm{G}^3_{s,k,1} \in \mathbb{R}^{S\times K\times 1}$ by the Tensor-Train decomposition as:
\begin{align}
\bm{W}_{i,j,k} = \sum_{r,s} \bm{G}^1_{1,i,r} \times \bm{G}^2_{r,j,s} \times \bm{G}^3_{s,k,1}.
\end{align}
The parameter compression ratio $\gamma_{\mbox{tt}}$ 
can be described as:
\begin{align}
\label{eq:tt_comp}
\gamma_{\mbox{tt}} = \frac{IR + RJS + SK}{IJK}.
\end{align}
% Tensor-Train decomposition can flexibly reduce parameters but cannot speed up inference as Tucker decomposition.
Tensor-Train decomposition provides flexible parameter reduction, as it involves $N-1$ hyperparameters corresponding to the rank of $N$-th order tensors. However, unlike Tucker decomposition, it cannot accelerate inference because the original tensor must be reconstructed during inference.

% On the other hand, Tensor-Train decomposition has the advantage of decomposition speed for large tensors.
% However its speed has little impact when the target is parameter reduction, since tensor decomposition is only performed when converting to a smaller model from middle model.

% \section{E2E SLU model}

\begin{figure}[t]
\centering
\vspace{-15pt}
\includegraphics[width=8.0cm]{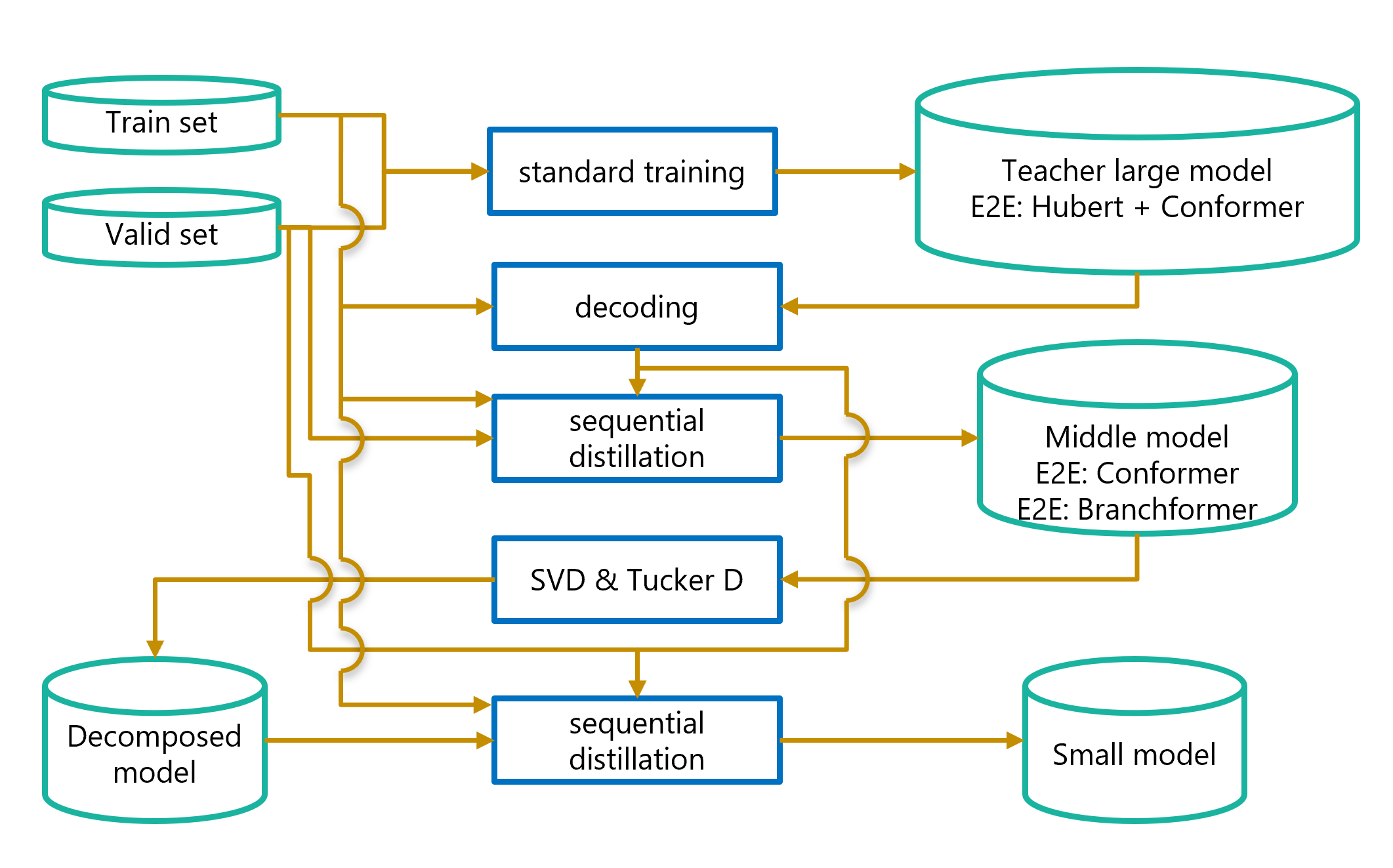}
\caption{Training procedure.}
\label{fig:overview}
\end{figure}

%\subsection{Decomposed E-Branchformer}
\section{Decomposed E2E SLU model}
\subsection{Training overview}
We investigate the tensor decomposition techniques in E2E SLU models with higher-order tensors as parameters.
E2E SLU is a task that estimates semantic label sequences $\bm{Y}$ directly from input speech feature sequences $\bm{X}$.
% Although the label is structural, we target the E2E model, which treats the label as a sequence of tokens for simplicity.
Fig.\ref{fig:overview} describes the overview of the training procedure of our model.
We use two-step sequential distillation to train a small model.
Recent SOTA models use self-supervised learning techniques, which have a large number of parameters (over 300M) \cite{hsu2021hubert}.
However, on-device processing requires a smaller model, typically around 30M or less.
Therefore, We do not apply tensor decomposition directly to the large model because the teacher model is too large to minimize within small parameters with sufficient rank.
We train small model through a middle-size model to achieve the target size with a compression ratio near 0.3.
%Therefore, to achieve the target size with a compression ratio of around 0.3, a middle-size model is trained before the small model.
As shown in the blue boxes in Fig.\ref{fig:overview}, the training procedure has 5 steps. 
% Please also refer to Fig.\ref{fig:overview}.
\begin{enumerate}
    \item A large teacher model is trained with ground truth labels. We use the E2E Conformer model with Hubert-based feature extractor\footnote{https://dl.fbaipublicfiles.com/hubert/hubert\_large\_ll60k.pt} \cite{hsu2021hubert}, which has achieved the best performance in prior study \cite{paden2023stop}.
    \item The training data is decoded using the trained teacher model to generate teacher labels.
    \item The middle model is trained with both the distilled and ground truth labels. The middle model is used as the seed model to miniaturize the final small model using the decomposition, so its structural type is the same as the final small model.
    \item Tensor decomposition and SVD techniques are applied to the middle model to reduce the parameters. We set the target size to 15M or 30M as in a previous study \cite{le2022deliberation}.
    
    \item Finally, the decomposed model is finetuned with distillation to generate the small model.
\end{enumerate}

% Although there are several candidates for the teacher model, 

%Using the teacher model, we reduce the parameter size from the large model by tensor decomposition, and sequential distillation techniques.

% based on E2E Conformer E-Branchformer (Enhanced Branchformer) \cite{kim2023branchformer} 

%We miniaturize Conformer-based and E-Branchformer-based middle E2E SLU model to realize small E2E SLU model for on-device processing.
The middle model and small model have Conformer or E-Branchformer encoders.
%Conformer is a variant of Transformer with added convolutional layers.
%In addition, E-Branchformer is enhanced Branchformer, whis is a parallelized version of the Conformer, rather than a sequential one.
%Conformer is a very powerful model, and further E-Branchformer has been reported to achieve higher accuracy than Conformer \cite{peng2022branchformer,kim2023branchformer}.
% We apply SVD to the linear layers and Tucker decomposition to the convolution layers in the models.
Conformer and E-Branchformer include convolution layers that have 3rd or 4th-order tensor parameters.
Thus, they can be efficiently compressed by tensor decomposition techniques such as Tucker decomposition.

\subsection{Sequential distillation}
Sequential distillation is used to downsize sequential models such as encoder-decoder models \cite{kim2016sequence}.
Frame-by-frame distillation is performed by the KL-divergence minimization criterion using softmax outputs with temperature hyperparameters.
In contrast, sequential distillation uses the labels obtained by inference of the teacher model as target labels for student model training.
It is also reported to be more effective when extended to N-best from 1-best \cite{mun2019sequence}.
However, in our setting, when extended to 5-best, the performance is slightly degraded.
Instead, our system uses sequential distillation, which combines the 1-best obtained by the teacher model and ground truth labels.

\subsection{Determination of rank from compression ratio}

In our model, the Tucker decomposition is mainly used to reduce the parameters of the convolution layer.
The reasons for adopting the Tucker decomposition are that it is expected to be able to represent the convolution weight tensor with fewer parameters than CP decomposition and that it contributes not only to parameter reduction but also to speed up \cite{kim2015compression}.
The tucker decomposition can be inferred without reconstructing the original tensor by swapping the order of computation as in SVD.
This technique is very compatible with on-device processing.
On the other hand, Tensor-Train decomposition cannot accelerate inference because the original tensor must be reconstructed during inference.

The rank of the decompositions is determined by the specified compression ratio.
We apply SVD to the linear layers and Tucker decomposition to the convolution layers in the models.
The number of nodes in the middle of the SVD and the size of the core tensor of the Tucker decomposition are determined from the given compression ratio (eq. (\ref{eq:svd_comp})).
In the case of SVD, the number of middle nodes $R$ can be calculated as follows:
\begin{align}
    R = \gamma_{\mbox{svd}} \frac{IJ}{I + J}.
\end{align}
%In the case of SVD, it can be uniquely　calculated from the compression ratio.
On the other hand, since the core tensor shape cannot be uniquely determined from the given compression ratio (eq. (\ref{eq:tucker_comp})), the rank of the Tucker decomposition is calculated by iteratively halving all dimensions until the current ratio is under the given one as in Algorithm 1.
In line 4, all dimensions are halved, and each time the compression ratio is recalculated in line 14.
Lines 5--13 compensate so that each dimension is not less than 1.

%For comparison, CP decomposition and Tensor-Train decomposition is adopted instead of the Tucker decomposition.
For comparison, the Tucker decomposition is replaced by CP and Tensor-Train decomposition.
CP decomposition, similar to SVD, can uniquely determine rank from compression ratio (eq. (\ref{eq:cp_comp})).
However, Tensor-Train decomposition cannot be uniquely determined from the given compression ratio (eq. (\ref{eq:tt_comp})).
Therefore, the dimension is iteratively reduced, as Tucker decomposition.

%Note that although Kim proposed speeding up technique to compute the Tucker decomposition efficiently by restricting the decomposition to channels\cite{kim2015compression}, we applied the decomposition to all dimensions because inference speed is not included in the evaluation criteria of STOP challenge.

\section{Experimental evaluation}
\subsection{Experimental settings}
We evaluated our system on the STOP dataset \cite{paden2023stop}.
STOP dataset consists of over 200,000 audio files from over 800 speakers and text and semantic parses.
They are divided into train, valid, and test sets.
Evaluation criteria were performed by exact match accuracy (EM), which is the percentage of perfect match of the label sequences.

The teacher model was based on the Conformer model.
The model used a HuBERT-based feature extractor and convolution layers to reduce the input feature length.
The encoder had 12 attention blocks, each with 512 dimensions with 8 heads.
The decoder had 6 attention blocks, and each block also had 512 dimensions with 8 heads.
The number of parameters of the teacher model was 431M.
%430,635,985.
The Conformer-based middle model had 47M parameters.
%47,443,048 parameters. 
Its encoder had 10 attention blocks, and each block had 384 dimensions with 6 attention heads.
The decoder had 3 blocks with the same dimensions and attention heads.
On the other hand, the E-Branchformer-based middle model had 53M parameters.
%53,048,376 parameters.
Its encoder had 10 attention blocks, each with 384 dimensions with 6 heads.
The decoder had 3 blocks with the same dimensions and heads.

For the training of the middle and small models, we used speed perturbation.
%In the sequential distillation, the large model which had the same encoder type (Conformer/Branchformer) was used for teacher model.
%The distilled labels and ground truth labels were both used and they were both expanded with speed perturbation for the distillation.
The teacher and ground truth labels were both used with speed perturbation for the distillation.
On the other hand, the validation data consisted of ground truth text, even in the sequential distillation.
The input feature was log mel-filter bank having 80 bins.
Furthermore, we used SpecAugment technique \cite{park2019specaugment}.
After that, we applied utterance-level mean normalization.
The target semantic parse labels were divided using the byte-pair encoding (BPE) model with 500 tokens.
Adam optimization with warmup scheduling was used in our training.
Finally, we used an averaged model of the top 10 accuracy checkpoints.

\subsection{Comparison of Conformer and E-Branchformer}
Table \ref{table:results_15M} shows the experimental results of our system on the STOP dataset with 15M parameter limitation.
In this experiment, we used tucker decomposition (Sec \ref{sec:tucker}) to reduce the convolution parameters.
%The simple small E2E Conformer model\cite{gulati2020conformer} which was trained from scratch was quite worse than the large model.
%However, b
For tensor decomposition, we set the compression ratio of the encoder and decoder as 0.3 and 0.295, respectively, in Conformer.
In the case of the E-Branchformer, the compression ratios were set to 0.25 and 0.3 for the encoder and decoder.
By using tensor decompositions from the middle Conformer, the small Conformer achieved 65.4 EM.
Furthermore, the E-Branchformer encoder significantly improved the performance.
\begin{comment}
Branchformer has reported better performance than Conformer with more than 30M parameters \cite{peng2022branchformer}.
Track 2 is limited to 15M or less, but by applying a tensor decomposition with a middle model ($>30\mbox{M}$), E-Branchformer maintained the trend with a small number of parameters.
\end{comment}
Our system, E-Branchformer-based E2E SLU, achieved 70.1\% EM for the test set with a parameter count of 15M, which was better than the previous study \cite{le2022deliberation}.
This result was a higher performance than the teacher model (69.4\%).
This was due to that the teacher model was based on a Conformer encoder and ground truth labels were also used for sequential distillation.
Furthermore, under this condition, it appeared that the small model had not fully converged, so we continued training it for an additional 200 epochs.
As a result, the accuracy was increased to 70.9\%.

In addition we made comparisons with 30M parameters.
Table \ref{table:results_30M} shows the experimental results of our system on the STOP dataset with 30M parameter limitation.
In this limitation, we set the compression ratios of the encoder and decoder as 0.65 and 0.65 in Conformer.
In the E-Branchformer, the encoder and decoder compression ratios were set as 0.55 and 0.6, respectively.
The performance of the E-Branchformer was higher than Conformer but lower than the deliberation model \cite{le2022deliberation}.
The comparison with 15M indicates that the E2E model is more advantageous when constructing the smaller model.
%Our E2E model does not explicitly separate ASR and NLU, so the parameters can be reduced in a balanced way so that SLU performance becomes higher.
Our E2E model does not explicitly separate ASR and NLU, so unlike \cite{le2022deliberation}, we can reduce the parameters in a balanced manner so that SLU performance remains high.

%Prior work \cite{le2022deliberation} compares 30M and 15M models, but these both models have same 5M parameter NLU.
% Thus, only the ASR parameter is reduced by 15M. 
% This parameter reduction is unbalanced, and the performance has been degra ded significantly. 
%On the other hand, our E2E model does not explicitly separate ASR and NLU, so the parameters can be reduced in a balanced way so that SLU performance becomes higher.

\begin{figure}[t]
\centering
\includegraphics[width=8.0cm]{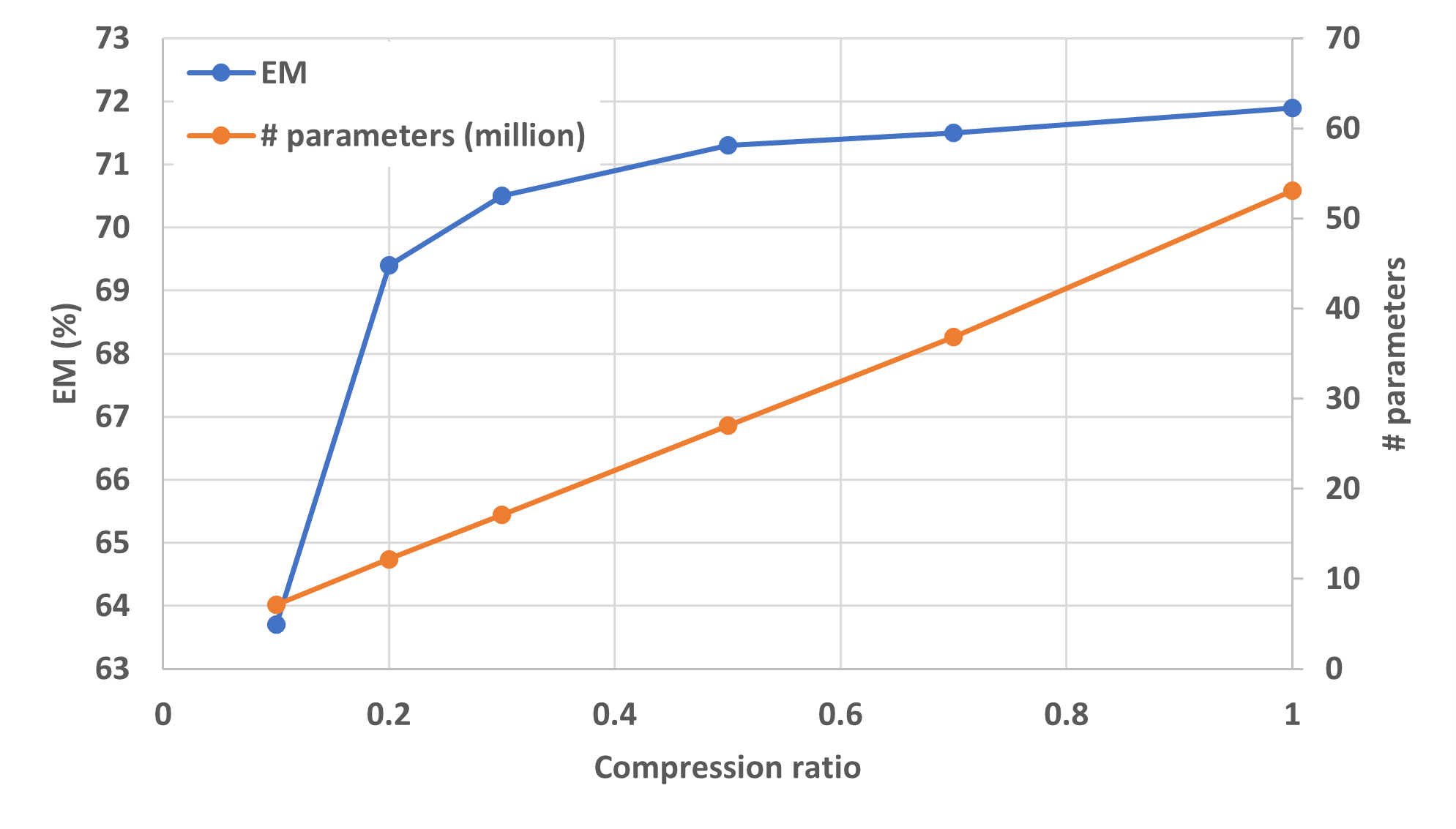}
\caption{EM of our system varying compression ratio with Tucker decomposition.}
\label{fig:varying_compression_ratio}
\end{figure}

\begin{table}[t]
 \caption{Experimental results (token accuracy and EM) on STOP dataset within 15M changing decomposition type.}
 \label{table:results_decomposed_type}
 \centering
  \begin{tabular}{lcc}
    \hline
           & TAcc. & EM \\
    \hline
    Tucker decomposition (Sec. \ref{sec:tucker}) & {\bf 91.8 } & 70.1 \\
    CP     decomposition (Sec. \ref{sec:cp}) & 34.8 & 0.1\\
    Tensor-Train decomposition (Sec. \ref{sec:tt}) & 90.7 & {\bf 70.2 }\\
    \hline
  \end{tabular}
  \vspace{-8pt}
\end{table}

\subsection{Comparison of different compression ratios}
Consequently, we investigated the performance changes with various compression ratios.
Figure \ref{fig:varying_compression_ratio} shows the EM with varying compression ratio in E-Branchformer with Tucker decomposition.
We used the same compression ratio for both encoder and decoder in this experiment.
According to the figure, 
The EM drops sharply when the compression ratio falls under 0.3.
Therefore, it is important to set an appropriate compression ratio when applying tensor decomposition.
% On the other hand, the number of parameters drops almost linearly with the compression ratio, indicating that the overhead has little effect.
%Thus, it was reasonable to determine the size of the source model for the tensor decomposition so that the compression ratio around 0.3 was the target size.

\subsection{Comparison of different tensor decompositions}
We further examined the performance with different tensor decomposition techniques.
In addition to the Tucker decomposition used in our system, we applied CP decomposition and Tensor-Train decomposition to the convolution layer.
Table \ref{table:results_decomposed_type} shows the performance on the STOP dataset.
In this experiment, the compression ratio was set to 0.25 and 0.3 for the encoder and decoder, respectively, targeting 15M.
CP decomposition improved the training accuracy from the initial parameters, but it reached a saturation point at small epochs.
This was because the CP decomposition decomposed all dimensions with the same rank, which led to an excessive loss of expressiveness.
In contrast, the Tensor-Train decomposition performed significantly well, as did the Tucker decomposition.
EM was slightly better for the Tensor-Train decomposition, but we observed a large difference in token accuracy.

\begin{comment}
\begin{table}[t]
 \caption{Experimental results with fine tuning to 200 epochs.}
 \label{table:results_finetune200}
 \centering
  \begin{tabular}{rcc}
    \hline
           epochs & Valid/Test EM \\
    \hline
    50  & 69.6 / 70.1 \\
    100  & 70.1 / 70.7  \\
    150  & 70.3 / 70.8 \\
    200  & {\bf 70.4 / 70.9 } \\
    \hline
  \end{tabular}
\end{table}
\end{comment}

\begin{comment}

\subsection{Additional training}
Finally, we finetuned the Tucker decomposition-based E-Branchformer model up to 200 epochs, having stopped it at 50 epochs.
Figure \ref{fig:results_200epochs} shows the EM varying the training epochs.
It almost converged at 100 epochs, but the EM improved slightly after that.
Finally, we finished the training at 200 epochs and achieved a 70.9\% EM with 15M parameters.

\begin{figure}[t]
\centering
\includegraphics[width=8.0cm]{200epochs.png}
\caption{Experimental results with fine tuning to 200 epochs using Tucker decomposition.}
\label{fig:results_200epochs}
\end{figure}
\end{comment}

\section{Conclusion}
In this paper, we describe our investigation of the minimization of the E-Branchformer for on-device E2E SLU.
We applied sequential distillation and tensor decomposition techniques to the E-Branchformer.
Compared to the Conformer-based model, E-Branchformer with Tucker decomposition achieved higher performance in both 15M and 30M limitations.
In addition, it obtained better performance than the deliberation model in 15M limitation.
The experiment with different compression ratios showed that compression around 0.3 points was efficient for E-Branchformer.
Our comparison of the Tucker decomposition with the CP decomposition and Tensor-Train decomposition showed that the Tucker decomposition was a relatively efficient decomposition.
Finally, our system, E-Branchformer-based decomposed E2E SLU model, achieved 70.9\% EM with 15M parameter limitation on STOP data.

%, which is a disadvantage of the conventional E-Branchformer.

%{ \small
\begin{comment}
\section{Acknowledgement}
This work used Bridges2 system at PSC and Delta system at NCSA through allocation CIS210014 from the Advanced Cyberinfrastructure Coordination Ecosystem: Services \& Support (ACCESS) program, which is supported by National Science Foundation grants \#2138259, \#2138286, \#2138307, \#2137603, and \#2138296.

J.H. was supported by the NSF Graduate Research Fellowship under Grant Nos. DGE1745016 and DGE2140739. The opinions expressed in this paper do not necessarily reflect those of that funding agency.
%}
\end{comment}

\newpage

\bibliographystyle{IEEEbib}
% \footnotesize
{ %\footnotesize%\small
\bibliography{refs}

\begin{thebibliography}{10}

\bibitem{serdyuk2018towards}
Dmitriy Serdyuk, Yongqiang Wang, Christian Fuegen, Anuj Kumar, Baiyang Liu, and
  Yoshua Bengio,
\newblock ``Towards end-to-end spoken language understanding,''
\newblock in {\em 2018 IEEE International Conference on Acoustics, Speech and
  Signal Processing (ICASSP)}. IEEE, 2018, pp. 5754--5758.

\bibitem{radfar2020end}
Martin Radfar, Athanasios Mouchtaris, and Siegfried Kunzmann,
\newblock ``End-to-end neural transformer based spoken language
  understanding,''
\newblock {\em Proc. Interspeech 2020}, pp. 866--870, 2020.

\bibitem{dong2022improving}
Jingjing Dong, Jiayi Fu, Peng Zhou, Hao Li, and Xiaorui Wang,
\newblock ``Improving spoken language understanding with cross-modal
  contrastive learning,''
\newblock {\em Proc. Interspeech 2022}, pp. 2693--2697, 2022.

\bibitem{wang2022adversarial}
Ye~Wang, Baishun Ling, Yanmeng Wang, Junhao Xue, Shaojun Wang, and Jing Xiao,
\newblock ``Adversarial knowledge distillation for robust spoken language
  understanding,''
\newblock {\em Proc. Interspeech 2022}, pp. 2708--2712, 2022.

\bibitem{arora2022two}
Siddhant Arora, Siddharth Dalmia, Xuankai Chang, Brian Yan, Alan Black, and
  Shinji Watanabe,
\newblock ``Two-pass low latency end-to-end spoken language understanding,''
\newblock {\em Proc. Interspeech 2022}, pp. 3478--3482, 2022.

\bibitem{radfar2021fans}
Martin Radfar, Athanasios Mouchtaris, Siegfried Kunzmann, and Ariya Rastrow,
\newblock ``{FANS}: Fusing {ASR} and {NLU} for on-device {SLU},''
\newblock in {\em Interspeech 2021}, 2021.

\bibitem{le2022deliberation}
Duc Le, Akshat Shrivastava, Paden~D. Tomasello, Suyoun Kim, Aleksandr Livshits,
  Ozlem Kalinli, and Michael~L. Seltzer,
\newblock ``Deliberation model for on-device spoken language understanding,''
\newblock in {\em Interspeech 2022}, 2022, pp. 3468--3472.

\bibitem{saade2019spoken}
Alaa Saade, Joseph Dureau, David Leroy, Francesco Caltagirone, Alice Coucke,
  Adrien Ball, Cl{\'e}ment Doumouro, Thibaut Lavril, Alexandre Caulier,
  Th{\'e}odore Bluche, et~al.,
\newblock ``Spoken language understanding on the edge,''
\newblock in {\em 2019 Fifth Workshop on Energy Efficient Machine Learning and
  Cognitive Computing-NeurIPS Edition (EMC2-NIPS)}. IEEE, 2019, pp. 57--61.

\bibitem{tyagi2020fast}
Akshit Tyagi, Varun Sharma, Rahul Gupta, Lynn Samson, Nan Zhuang, Zihang Wang,
  and Bill Campbell,
\newblock ``Fast intent classification for spoken language understanding
  systems,''
\newblock in {\em ICASSP 2020-2020 IEEE International Conference on Acoustics,
  Speech and Signal Processing (ICASSP)}. IEEE, 2020, pp. 8119--8123.

\bibitem{desot2022end}
Thierry Desot, Fran{\c{c}}ois Portet, and Michel Vacher,
\newblock ``End-to-end spoken language understanding: Performance analyses of a
  voice command task in a low resource setting,''
\newblock {\em Computer Speech \& Language}, vol. 75, pp. 101369, 2022.

\bibitem{gupta-2022-building}
Akshat Gupta,
\newblock ``On building spoken language understanding systems for low resourced
  languages,''
\newblock in {\em Proceedings of the 19th SIGMORPHON Workshop on Computational
  Research in Phonetics, Phonology, and Morphology}, Seattle, Washington, July
  2022, pp. 1--11, Association for Computational Linguistics.

\bibitem{wang2022bottleneck}
Pu~Wang et~al.,
\newblock ``Bottleneck low-rank transformers for low-resource spoken language
  understanding,''
\newblock {\em Proceedings Interspeech 2022}, 2022.

\bibitem{avila2022low}
Anderson~R Avila, Khalil Bibi, Rui~Heng Yang, Xinlin Li, Chao Xing, and Xiao
  Chen,
\newblock ``Low-bit shift network for end-to-end spoken language
  understanding,''
\newblock {\em Proc. Interspeech 2022}, pp. 2698--2702, 2022.

\bibitem{gao2022meta}
Yingying Gao, Junlan Feng, Chao Deng, and Shilei Zhang,
\newblock ``Meta auxiliary learning for low-resource spoken language
  understanding,''
\newblock {\em Proc. Interspeech 2022}, pp. 2703--2707, 2022.

\bibitem{dinarelli2022toward}
Marco Dinarelli, Marco Naguib, and Fran{\c{c}}ois Portet,
\newblock ``Toward low-cost end-to-end spoken language understanding,''
\newblock {\em Proc. Interspeech 2022}, pp. 2728--2732, 2022.

\bibitem{gulati2020conformer}
Anmol Gulati, James Qin, Chung-Cheng Chiu, Niki Parmar, Yu~Zhang, Jiahui Yu,
  Wei Han, Shibo Wang, Zhengdong Zhang, Yonghui Wu, et~al.,
\newblock ``Conformer: Convolution-augmented transformer for speech
  recognition,''
\newblock {\em Proc. Interspeech 2020}, pp. 5036--5040, 2020.

\bibitem{peng2022branchformer}
Yifan Peng, Siddharth Dalmia, Ian Lane, and Shinji Watanabe,
\newblock ``Branchformer: Parallel mlp-attention architectures to capture local
  and global context for speech recognition and understanding,''
\newblock in {\em International Conference on Machine Learning}. PMLR, 2022,
  pp. 17627--17643.

\bibitem{kim2023branchformer}
Kwangyoun Kim, Felix Wu, Yifan Peng, Jing Pan, Prashant Sridhar, Kyu~J Han, and
  Shinji Watanabe,
\newblock ``E-{B}ranchformer: Branchformer with enhanced merging for speech
  recognition,''
\newblock in {\em 2022 IEEE Spoken Language Technology Workshop (SLT)}. IEEE,
  2023, pp. 84--91.

\bibitem{michael2021comparison}
Jeffrey~Josanne Michael, Nagendra~Kumar Goel, Jonas Robertson, Shravan Mishra,
  et~al.,
\newblock ``Comparison of svd and factorized tdnn approaches for speech to
  text,''
\newblock {\em arXiv preprint arXiv:2110.07027}, 2021.

\bibitem{zhang2021tiny}
Yuekai Zhang, Sining Sun, and Long Ma,
\newblock ``Tiny transducer: A highly-efficient speech recognition model on
  edge devices,''
\newblock in {\em ICASSP 2021-2021 IEEE International Conference on Acoustics,
  Speech and Signal Processing (ICASSP)}. IEEE, 2021, pp. 6024--6028.

\bibitem{denton2014exploiting}
Emily~L Denton, Wojciech Zaremba, Joan Bruna, Yann LeCun, and Rob Fergus,
\newblock ``Exploiting linear structure within convolutional networks for
  efficient evaluation,''
\newblock {\em Advances in neural information processing systems}, vol. 27,
  2014.

\bibitem{arora2022espnet}
Siddhant Arora, Siddharth Dalmia, Pavel Denisov, Xuankai Chang, Yushi Ueda,
  Yifan Peng, Yuekai Zhang, Sujay Kumar, Karthik Ganesan, Brian Yan, et~al.,
\newblock ``Espnet-slu: Advancing spoken language understanding through
  espnet,''
\newblock in {\em ICASSP 2022-2022 IEEE International Conference on Acoustics,
  Speech and Signal Processing (ICASSP)}. IEEE, 2022, pp. 7167--7171.

\bibitem{tjandra2017compressing}
Andros Tjandra, Sakriani Sakti, and Satoshi Nakamura,
\newblock ``Compressing recurrent neural network with tensor train,''
\newblock in {\em 2017 International Joint Conference on Neural Networks
  (IJCNN)}. IEEE, 2017, pp. 4451--4458.

\bibitem{tucker1966some}
Ledyard~R Tucker,
\newblock ``Some mathematical notes on three-mode factor analysis,''
\newblock {\em Psychometrika}, vol. 31, no. 3, pp. 279--311, 1966.

\bibitem{kim2015compression}
Yong-Deok Kim, Eunhyeok Park, Sungjoo Yoo, Taelim Choi, Lu~Yang, and Dongjun
  Shin,
\newblock ``Compression of deep convolutional neural networks for fast and low
  power mobile applications,''
\newblock {\em arXiv preprint arXiv:1511.06530}, 2015.

\bibitem{harshman1970foundations}
Richard~A Harshman et~al.,
\newblock ``Foundations of the parafac procedure: Models and conditions for an
  "explanatory" multimodal factor analysis,''
\newblock {\em UCLA working papers in phonetics}, 1970.

\bibitem{oseledets2011tensor}
Ivan~V Oseledets,
\newblock ``Tensor-{T}rain decomposition,''
\newblock {\em SIAM Journal on Scientific Computing}, vol. 33, no. 5, pp.
  2295--2317, 2011.

\bibitem{hsu2021hubert}
Wei-Ning Hsu, Benjamin Bolte, Yao-Hung~Hubert Tsai, Kushal Lakhotia, Ruslan
  Salakhutdinov, and Abdelrahman Mohamed,
\newblock ``{HuBERT}: Self-supervised speech representation learning by masked
  prediction of hidden units,''
\newblock {\em IEEE/ACM Transactions on Audio, Speech, and Language
  Processing}, vol. 29, pp. 3451--3460, 2021.

\bibitem{paden2023stop}
Paden Tomasello, Akshat Shrivastava, Daniel Lazar, Po-Chun Hsu, Duc Le, Adithya
  Sagar, Ali Elkahky, Jade Copet, Wei-Ning Hsu, Yossi Adi, et~al.,
\newblock ``{STOP}: A dataset for spoken task oriented semantic parsing,''
\newblock in {\em 2022 IEEE Spoken Language Technology Workshop (SLT)}. IEEE,
  2023, pp. 991--998.

\bibitem{kim2016sequence}
Yoon Kim and Alexander~M Rush,
\newblock ``Sequence-level knowledge distillation,''
\newblock in {\em Proceedings of the 2016 Conference on Empirical Methods in
  Natural Language Processing}, 2016, pp. 1317--1327.

\bibitem{mun2019sequence}
Raden~Mu’az Mun’im, Nakamasa Inoue, and Koichi Shinoda,
\newblock ``Sequence-level knowledge distillation for model compression of
  attention-based sequence-to-sequence speech recognition,''
\newblock in {\em ICASSP 2019-2019 IEEE International Conference on Acoustics,
  Speech and Signal Processing (ICASSP)}. IEEE, 2019, pp. 6151--6155.

\bibitem{park2019specaugment}
Daniel~S Park, William Chan, Yu~Zhang, Chung-Cheng Chiu, Barret Zoph, Ekin~D
  Cubuk, and Quoc~V Le,
\newblock ``{SpecAugment}: A simple data augmentation method for automatic
  speech recognition,''
\newblock {\em Proc. Interspeech 2019}, pp. 2613--2617, 2019.

\end{thebibliography}
}

\end{document}